%% file: main.tex
\documentclass{IOS-Book-Article}

\usepackage{tabularx}
\usepackage{url}
\usepackage{booktabs}
\usepackage{lscape}
\usepackage{longtable}
\usepackage[T1]{fontenc}
\usepackage{tikz}
\usepackage{hyperref}

\usepackage{pdflscape}
\usepackage{tabularx}
\usepackage{amsmath}
\usepackage{array}
\usepackage{booktabs}
\usepackage{makecell}
\usepackage{multirow}
\usepackage{caption}
\usepackage{graphbox}

\usepackage{amsthm}

\newcommand{\atleast}[1]{{\geqslant}{#1}}

\usepackage{mathptmx}
\usepackage{newtxmath}
\usepackage[T1]{fontenc}
\usepackage{soul}\setuldepth{article}
\def\hb{\hbox to 11.5 cm{}}

\begin{document}

\pagestyle{headings}
\def\thepage{}
\begin{frontmatter}              

\title{MAECO-Lite: Modular Ontology for Dynamic Malware Analysis}
\runningtitle{MAECO-Lite: Modular Ontology for Dynamic Malware Analysis}

\author[A]{\fnms{Zekeri} \snm{Adams}\orcid{0009-0006-3413-2409}%
\thanks{Corresponding Author: Zekeri Adams, adams1@uniba.sk.}},
\author[B]{\fnms{Peter} \snm{Švec}\orcid{0000-0002-8315-5301}},
\author[A]{\fnms{Ján} \snm{Kľuka}\orcid{0000-0002-3406-3574}},
\author[B]{\fnms{Roderik} \snm{Ploszek}\orcid{0000-0003-0634-9476}},
\author[A]{\fnms{Monday} \snm{Onoja}\orcid{0000-0003-2119-170X}},
\author[B]{\fnms{Štefan} \snm{Balogh}\orcid{000-0003-0634-9476}}
and
\author[A]{\fnms{Martin} \snm{Homola}\orcid{0000-0001-6384-9771}}

\runningauthor{Z. Adams et al.}
\address[A]{Department of Applied Informatics, Comenius University in Bratislava,\\
Mlynská dolina, 842~48 Bratislava, Slovakia}
\address[B]{Institute of Computer Science and Mathematics,
Faculty of Electrical Engineering and Information Technology,
Slovak University of Technology, Ilkovičova~3, Bratislava, Slovakia}

\begin{abstract}
Capturing dynamic malware behavior in a practical but still semantically precise manner remains a significant challenge in cyber threat intelligence. While standards
such as MAEC and STIX provide widely adopted vocabularies for describing malware artifacts and observations, they represent data with considerable complexity in structures that often obscure important ontological distinctions.  In particular, they tend to conflate enduring malware artifacts with the events generated during execution, thereby flattening distinctions that are central in foundational standards for ontology design. In this paper, we conduct a foundational ontological analysis of core MAEC and STIX constructs relevant to dynamic malware analysis relying on Unified Foundational Ontology (UFO) as a theoretical lens. Our analysis reveals some ontological mismatches arising from the conflation of artifacts, dispositions, and runtime events in MAEC and STIX that complicate coherent representation of dynamic malware behavior and, from a practical perspective, limit the ability to reason about execution traces.  Based on these insights, we propose MAECO-Lite, a lightweight ontology designed to represent data and operationalize their processing for dynamic malware analysis. The ontology adopts a modular structure centered on samples, processes, actions, system artifacts, and MITRE ATT\&CK Techniques, while maintaining a clear separation between enduring entities and runtime events. An initial evaluation using description logic concept learning algorithms shows that the simplified ontology significantly improves learning performance, demonstrating that ontologically grounded modelling can enhance both semantic clarity and computational usability.

\end{abstract}

\end{frontmatter}

\section{Introduction}

Malware, short for malicious software, refers to any program or code intentionally designed to disrupt, damage, or gain unauthorized access to computer systems \cite{kramer2010general}. It includes a wide range of forms such as viruses, worms, Trojans, ransomware, spyware, and rootkits, each pursuing distinct malicious objectives \cite{kramer2010general,10051488}. Cybercriminals deploy malware for purposes including espionage, data theft, financial gain, sabotage, and large-scale system disruption \cite{Patsakis2025,10.33736/jcsi.4739.2022}. In response to these challenges, cybersecurity experts have increasingly explored ontology-based techniques for malware detection and analysis \cite{chiang2010,008abs-2403-11669,007BaloghG23}. Ontologies provide a consistent and machine-interpretable way to represent malware actions, behaviors, and interactions among system objects, and couple them with constraints and rules, with the potential to enhance detection accuracy and explainability \cite{008abs-2403-11669}, thereby supporting malware research, information sharing, inference, and automated reasoning.
Integrating malware ontologies into security frameworks enables to:
\begin{itemize}
\item enhance threat intelligence sharing \cite{10.1145/3458027},
\item standardize malware classification and reporting \cite{jsss.2023.50}, and
\item support automated reasoning for detection and mitigation \cite{fi16030069}.
\end{itemize}
Existing malware ontologies mostly cover static malware analysis, which extracts features from binaries without execution. This has inherent limitations, particularly vulnerability to obfuscation \cite{MOLINACORONADO2025104094}, evasion tactics \cite{GENG2024103595}, and a limited ability to capture actual malware behavior \cite{static_limit,006ChowdhuryB22}. These shortcomings have motivated increased reliance on dynamic malware analysis, which observes malware behavior at runtime and yields richer behavioral evidence.
Dynamic malware analysis produces heterogeneous and complex data and imposes challenges for ontological representation: ranging from low-level host and network artifacts (e.g., files, processes, registry keys, and network flows) to higher-level behavioral abstractions such as runtime actions, composite behaviors, and malware capabilities. Transforming such heterogeneous evidence into actionable and explainable malware intelligence requires principled modelling frameworks that clearly distinguish \emph{what exists}, \emph{what happens}, and \emph{what is observed}. Such distinctions are essential for sound reasoning, consistent knowledge sharing, and trustworthy explanations. 

We study a widely adopted standard that captures both static and dynamic malware features, namely MAEC \cite{maec_core_concepts,maec_vocabularies} (Malware Attribute Enumeration and Characterization), by analysing its representational commitments through the lens of a well-established foundational ontology, UFO \cite{guizzardi2013,guizzardi2005} (Unified Foundational Ontology).
The objectives of this research are as follows:
\begin{enumerate}
    \item Conduct a foundational ontological analysis of MAEC using UFO to clarify their conceptual structures and relationships.
    \item Develop an ontology for dynamic malware analysis with particular focus on practicality and expressibility of explanations.
\end{enumerate}

\looseness=-1
Following these goals, we present a UFO-based ontological analysis of the MAEC standard, including the referenced STIX Cyber Observable Objects (SCOs) \cite{stix} denoting runtime artifacts and MITRE ATT\&CK \cite{mitre-attack} techniques capturing action realizations.
The analysis provides a principled grounding for malware modelling, explicitly distinguishing between malware artifacts (endurants), their capabilities (dispositions), and runtime behaviors (perdurants), and model actions as temporally extended events involving well-defined participants. It thus exposes the ambiguities in MAEC’s representation, and enables to clarify the conceptual structure and relationships among its constructs. Building on this analysis, we introduce MAECO-Lite, an ontology for practical computational analysis of dynamic malware behavior, aligned with UFO. It centres on key MAEC constructs, adopts a lightweight, modular design for tractability, and incorporates relevant STIX SCOs and MITRE ATT\&CK Techniques. We also show an initial evaluation using description logic concept learning algorithms, demonstrating that compared to a naive baseline, the ontology significantly improves learning performance, and that ontologically grounded modelling can enhance both semantic clarity and computational usability.

\section{Preliminaries}

\subsection{MAEC and STIX Overview}

The Malware Attribute Enumeration and Characterization (MAEC) language~\cite{maec_core_concepts, maec_vocabularies} is a structured, machine‑readable schema designed to represent information produced during malware analysis. Developed and maintained by the MITRE Corporation, MAEC provides a standardized mechanism for encoding malware samples, their attributes, observed behaviors, and resulting effects in a consistent and interoperable manner. By formalizing the representation of malware analysis results, MAEC aims to facilitate information sharing, comparison, and automated processing across tools, organizations, and analysis environments. MAEC content is typically serialized in platform‑independent formats such as JSON or XML, enabling seamless integration into malware analysis pipelines and threat intelligence workflows. In practice, MAEC is supported by widely used dynamic malware analysis platforms, such as Cuckoo Sandbox, which can generate MAEC‑compliant reports directly from sandbox executions~\cite{cuckoo2012reporting}.

MAEC organizes malware knowledge around a set of top‑level objects that capture core concepts relevant to malware characterization.
Figure 1 illustrates malware characterization using the MAEC 5.0 standard \cite{maec_core_concepts}, showing MAEC top-level objects and their relationships, including links to STIX observables. 

\begin{figure}[tb]
\centering
\includegraphics[width=0.9\textwidth]{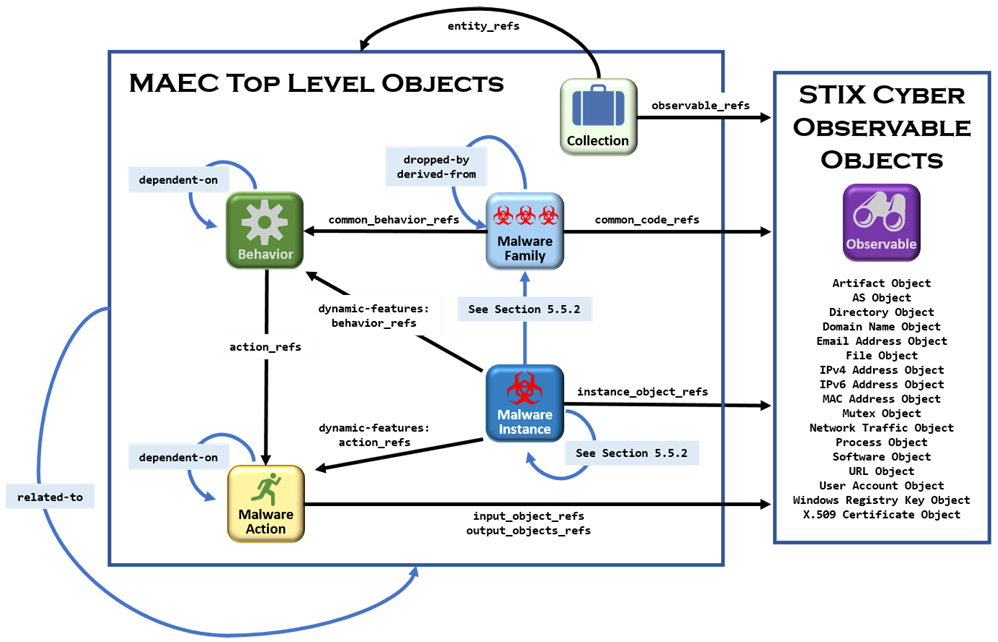}
\caption{Overview of MAEC-STIX \cite{maec_core_concepts}}
\label{fig:maec-stix}
\end{figure}
Central to the model are \emph{malware instances} and \emph{malware families}, which support the identification, classification, and grouping of related samples. To describe runtime activity, MAEC distinguishes between \emph{actions} and \emph{behaviors}. \emph{Malware actions} represent low‑level operational steps, such as API calls or system interactions observed during execution, while behaviors 
provide a higher‑level abstraction that groups related actions according to their purpose (for instance, the capture-keyboard-input behavior may implement the credentials harvesting capability, persist-after-system-reboot is a behavior implementing persistence, etc). MAEC also provides collection constructs to support the grouping of related analysis artifacts. A Collection captures a set of MAEC entities, such as malware instances, behaviors, and capabilities, as well as STIX Cyber Observable Objects that are associated in some meaningful way. Collections enable analysts to aggregate and correlate heterogeneous analysis results across different samples, tools, or executions, supporting comparison, reuse, and knowledge sharing.

MAEC further introduces a rich set of types that refine and specialize its top-level objects, thereby enabling a more fine-grained representation of malware characteristics and analysis results. These types include, among others, API calls, binary obfuscation techniques, static and dynamic features, capabilities, process tree nodes, analysis and signature metadata, field data, names, and development environment information. Together, these constructs allow MAEC to describe malware from multiple analytical perspectives, ranging from low-level program artifacts to higher-level behavioral and contextual information.
For instance, MAEC \emph{capabilities} are used to capture the high-level abilities that a malware instance possesses. These capabilities abstract over lower-level behaviors that implement specific functionalities of the malware. A capability therefore represents a generalized description of what the malware can accomplish (e.g., persistence or data exfiltration), while behaviors describe the concrete mechanisms through which these abilities are realized. This abstraction supports generalized malware characterization and facilitates comparison across malware samples and families.

A key strength of MAEC is its support for explicit inter‑object referencing, which allows malware instances, actions, behaviors, and capabilities to be semantically linked. This design enables analysts to associate high‑level behavioral characterizations with the concrete low‑level operations that realize them, as well as with the observable artifacts produced during execution. Such linking is essential for dynamic malware analysis, where understanding how individual system interactions contribute to broader malicious objectives is critical.
Observable artifacts referenced in MAEC, such as files, processes, registry keys, and network communications, are commonly described using the Structured Threat Information eXpression (STIX) language~\cite{stix}. STIX provides a complementary framework for representing and exchanging cyber threat intelligence, with Cyber‑observable Objects (SCOs) offering a standardized vocabulary for concrete host  and network‑level entities. By embedding or referencing STIX observables within MAEC content, analysts can jointly represent malware behavior and its manifestation in real execution environments. This integration allows MAEC to focus on the behavioral semantics of malware while delegating the detailed representation of observable artifacts to STIX. Furthermore, the integration of MITRE ATT\&CK \cite{mitre-attack}, with its tactics and techniques, into MAEC supports richer and more interoperable representations of malware intelligence derived from static, dynamic, or hybrid analysis.

\subsection{Unified Foundational Ontology (UFO)}

The Unified Foundational Ontology (UFO) \cite{guizzardi2013,guizzardi2005,guizzardi2022,Guizzardii2015} is a formal ontology framework designed to provide a principled foundation for conceptual modeling. It distinguishes between types (universals) and individuals, which are mutually disjoint. A defining characteristic of types is that they may have instances, and they can also have subtypes. UFO includes several categories of universals, among which are kinds, subkinds, and roles: kinds are rigid, essential types that all instances must always belong to (e.g., Vehicle); subkinds are specialized versions of kinds (e.g., Passenger Vehicle, Commercial Vehicle); roles are context-dependent types that instances may acquire or lose over time (e.g., Driver). A specific vehicle, such as a particular delivery van, is an individual instantiating the type Vehicle as a first-order universal \cite{Guizzardii2015,guizzardi2013}.

\emph{Individuals} are further classified into abstract and concrete individuals. \emph{Abstract individuals} include entities such as numbers, sets, and propositions, which do not exist in space and time. \emph{Concrete individuals}, by contrast, are spatiotemporal entities and are disjointly categorized into Objects (Edurants) and events (Perdurants). \emph{Endurants} are individuals that exist in time with all their parts. They have essential and accidental properties. \emph{Perdurants} are individuals that unfold in time accumulating temporal parts. They are manifestations of dispositions and only exist in the past.

Endurants are subdivided into substantials and moments \cite{guizzardi2013,guizzardi2022}. Substantials exist independently and bear properties and relations (e.g., a Person, a Software System). Moments are dependent entities that inhere in other individuals and include modes and qualities. Modes are particularized properties that inhere in a single bearer and may themselves bear further moments; they include dispositions (e.g., a system’s capability to detect malware, a server’s vulnerability) and externally dependent modes (e.g., a user’s trust in a system), which depend on another entity. Dispositions manifest under certain conditions (e.g., a program’s ability to execute a task). Qualities are measurable properties, such as a file’s size or a system’s response time, and may also inhere in other aspects (e.g., the accuracy of a detection capability) \cite{Guizzardii2015}.

Events (perdurants)  represent transitions between situations, capturing changes from one state of affairs to another \cite{guizzardi2013,benevides2019,baratella2022}. A situation is a configuration of entities and their properties at a given time \cite{guizzardi2005, guizzardi2013,benevides2019}. Events arise as manifestations of dispositions activated in specific situations. A recurring pattern can be identified: situations activate dispositions, which are manifested through events involving their bearers, thereby producing new situations. For example, a software system may have a vulnerability that, under certain conditions, is exploited in an attack, resulting in a compromised state. Events thus change states of affairs by affecting objects and their aspects. Unlike objects or situations, events cannot change while preserving their identity; they are temporally bounded entities and, once completed, exist only in the \cite{guizzardi2013,benevides2019,baratella2022}.

\section{Ontological Analysis of MAEC and STIX Cyber-Observable Object}
\label{sc:onto-analysis-of-maec-stix}

\begin{figure}[tbhp]
\centering
\begin{minipage} {0.99\textwidth}
    \centering
        \includegraphics[width=\textwidth]{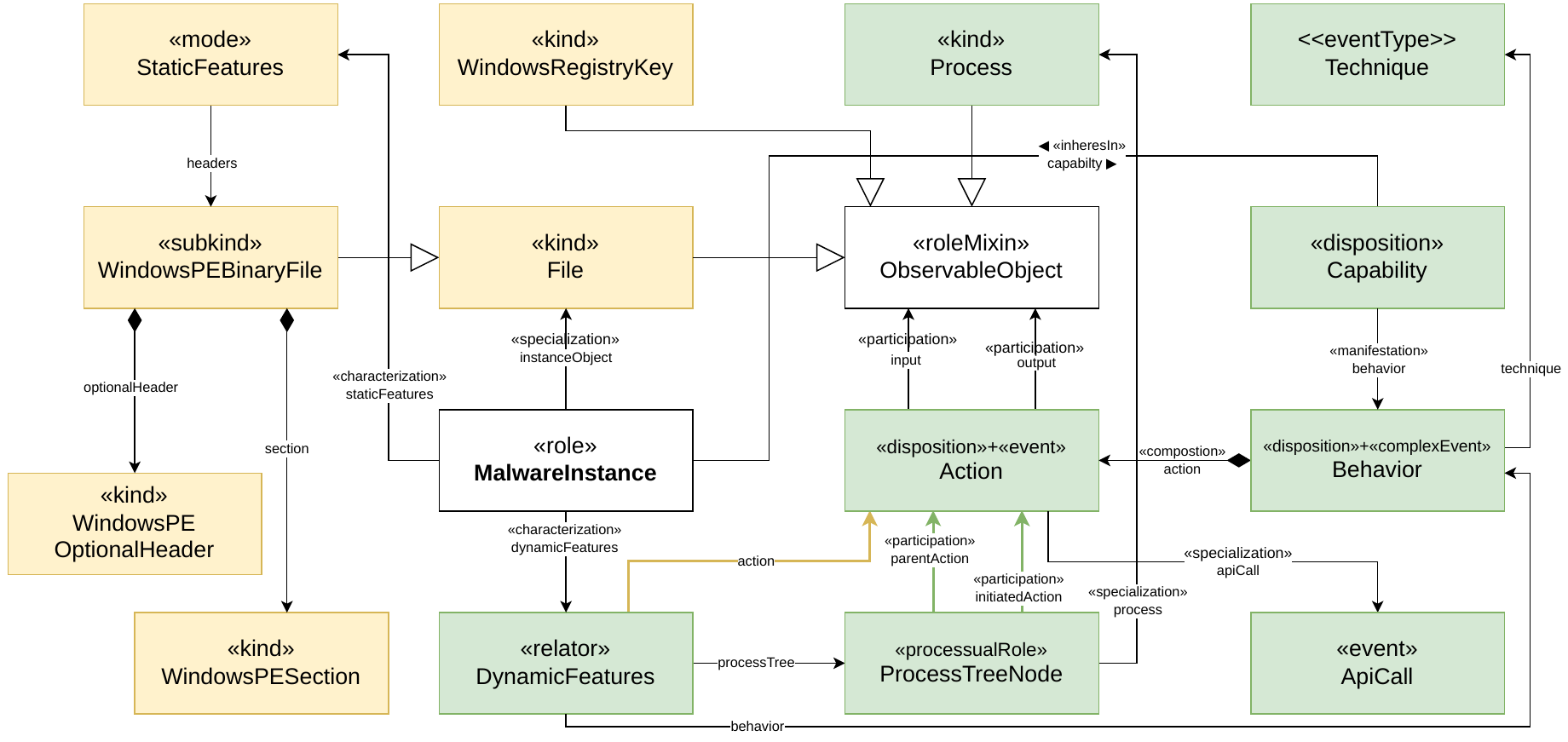}
    \caption{MAEC's Structure and UFO Representation }
    \label{fig:Ontology}
\end{minipage}
\end{figure}

Within MAEC’s top-level data model, the Malware Instance object occupies a structurally and semantically central role in representing malware in both static and dynamic analysis contexts. MAEC defines a \emph{malware instance} as a single member of a malware family, specified through its associated \emph{binary} file and enriched with static attributes, dynamic observations, capabilities, signatures, and analysis metadata. This design makes the \emph{malware instance} the focal point through which virtually all other MAEC constructs are linked. This includes: \emph{behaviors, malware actions, process trees, triggered signatures, analysis metadata}. As such, any ontology intending to model dynamic malware behavior must take the Malware Instance as its foundational anchor. The simplified UML diagram in Figure~\ref{fig:Ontology} illustrates the MAEC's Malware Instance type as the central anchoring concept of MAEC's concepts.

MAEC’s Malware Instance type is intended to denote a file under analysis; however, this is a misnomer, as both malicious and benign files may be analyzed. More precisely, it refers to a file whose malicious nature is determined through the analysis process. From a UFO perspective, Malware Instance is more appropriately understood as a role. In this view, a File assumes the role of a Malware Instance when subjected to malware analysis. The File type provides the identity criterion, while Malware Instance captures the contextual perspective under which the file is analyzed. So, Malware Instance should be modeled as a specialization of File, representing the role of a file under analysis.

MAEC's Malware Instance is mostly characterized by Static Features derived during static analysis and Dynamic Features obtained during dynamic analysis. Together, these features provide a comprehensive basis for analyzing and understanding the properties of the analyzed file. MAEC’s Malware Instance also bears one or more Capability instances, aligned to dispositions in UFO, inhering in the artifact. These dispositions represent the malware’s potential to manifest specific behaviors, such as keylogging or establishing a backdoor, under suitable execution conditions.

Static Features consist of binary and structural properties of a file (e.g., strings, certificates, and file headers), which MAEC captures as descriptors of its structural characteristics. From a UFO perspective, these features mostly correspond to intrinsic moments, as they inhere in the file and do not depend on external entities.

The MAEC Dynamic Features type captures features ``associated with the semantics of the executed code, of a malware instance'' \cite{maec_core_concepts}, mostly related to its runtime behavior. From a UFO perspective, it can be interpreted as a relator that inheres in the analyzed file and mediates its interactions with other entities (e.g., processes, files, or network resources) during execution. In this sense, it represents the behavioral context in which such interactions occur. It aggregates references to behaviors, actions, process structures, and network activity associated with execution. The referenced Behavior and Action entities correspond to events (perdurants) when they represent observed runtime phenomena. However, MAEC distinguishes between Actions obtained from static analysis (\texttt{action\_refs}) and those observed during execution (linked via a Process Tree Node's \texttt{initialed\_action} property). While the latter model actual event occurrences, the former model inferred or potential actions and are more appropriately interpreted as disposition. This collapses the distinction between events and dispositions. Moreover, a Process Tree Node relates a process to the actions it initiated, but also to the action that created it, and even indicates its position among other processes ordered by the time of creation. These aspects suggest mutually incompatible interpretations of this type in UFO: a relation (creation order) and two different kinds of processual roles: the subject of creation and the initiator of actions. Overall, the Dynamic Features type mixes entities, events, and their descriptions without explicit ontological separation, leading to ambiguous interpretation.

Furthermore, MAEC's Behavior is defined as “the specific purpose associated with a snippet of code as executed by a malware instance.” This description aligns Behavior with disposition in UFO, as it refers to what the malware is capable of doing. However, as analyzed in the context of Dynamic Features, Behaviors correspond to events. This dual interpretation mixes dispositions (endurants) with their manifestations as events (perdurants), thereby obscuring the ontological distinction between endurants and perdurants.

A Behavior can be linked to MITRE ATT\&CK Techniques ``used to implement'' it \cite{maec_core_concepts} and ATT\&CK documentation characterizes Techniques as methods of achieving certain malicious tactical goals \cite{mitre-attack}. The apparent intent of this link is to interpret clusters of Actions related to the Behavior as steps constituting the performance of the indicated Technique. However, a Technique is a general method that each malware may implement through its own pattern of Actions. These facts give rise to the following ontological interpretation: The actual observed Actions constitute a complex event that is an instance of the particular pattern of Actions implemented by the malware, which, in turn, is an instance of an ATT\&CK Technique. An individual Technique can thus be interpreted as an event universal in UFO.

\begin{table}[bt]
\centering
\caption{Ontological Alignment of MAEC 5.0 and STIX 2.1 Constructs with UFO}
\label{tab:maec-stix-ufo-mapping}
\resizebox{\textwidth}{!}{%
\input{tab-maec-stix-ufo-mapping}
}
\end{table}

An Action can be linked to STIX Cyber-observable Objects (SCOs) that participate in the event, either as “inputs” consumed by the Action or as “outputs” produced as a result of its execution \cite{maec_core_concepts}.
We interpret STIX Cyber-observable Objects not as ontological categories of entities, but as a unifying abstraction over the roles that heterogeneous entities (e.g., files, processes, network traffic) play in the context of observation and analysis. In this sense, ObservableObject is modeled as a role mixin in UFO: it is anti-rigid, as entities are only contingently observable; externally dependent, as observability arises from participation in analysis or monitoring events; and non-sortal, as it does not provide identity criteria but instead aggregates common relational properties across different roles. This interpretation allows observable entities to be consistently integrated into an event-based model, where they participate in events as inputs, outputs, or targets.

Table~\ref{tab:maec-stix-ufo-mapping} summarizes the alignment between MAEC 5.0 constructs and STIX 2.1 SCOs and the corresponding UFO categories.

\section{MAECO-Lite: Ontology Description and Alignment with UFO}

To provide a common foundation for ongoing research in the application of various explainable AI (XAI) methods to dynamic malware analysis, we have designed a lightweight ontology to semantically describe dynamic analysis reports of malware and benign samples.
The design goals include:
\begin{enumerate}
    \item Focus on the most relevant and easily interpretable behavioral characteristics derived from the sandbox execution of samples.
    \item Structural simplicity.
    \item Alignment with existing standards (MAEC, STIX, ATT\&CK) where applicable.
    \item Ontological alignment with UFO.
    \item Modularization, enabling selection of various coherent feature sets.
\end{enumerate}
The structural simplicity and modularization goals stem from the intended use in computational analysis. While methods such as DL concept learning \cite{XX10_vec2021ExperimentalEO} may work with ontologies directly, rich structures such as complex property chains pose increased computational demands. Many other machine learning methods are not able to process ontologies directly at all and require vectorization, where such rich structures pose even greater challenges. Modularization also facilitates focusing on ontologically coherent subsets of features or comparing the efficacy of malware detection based on different feature sets.

\begin{figure}[tbp]
\centering
\includegraphics[width=\textwidth]{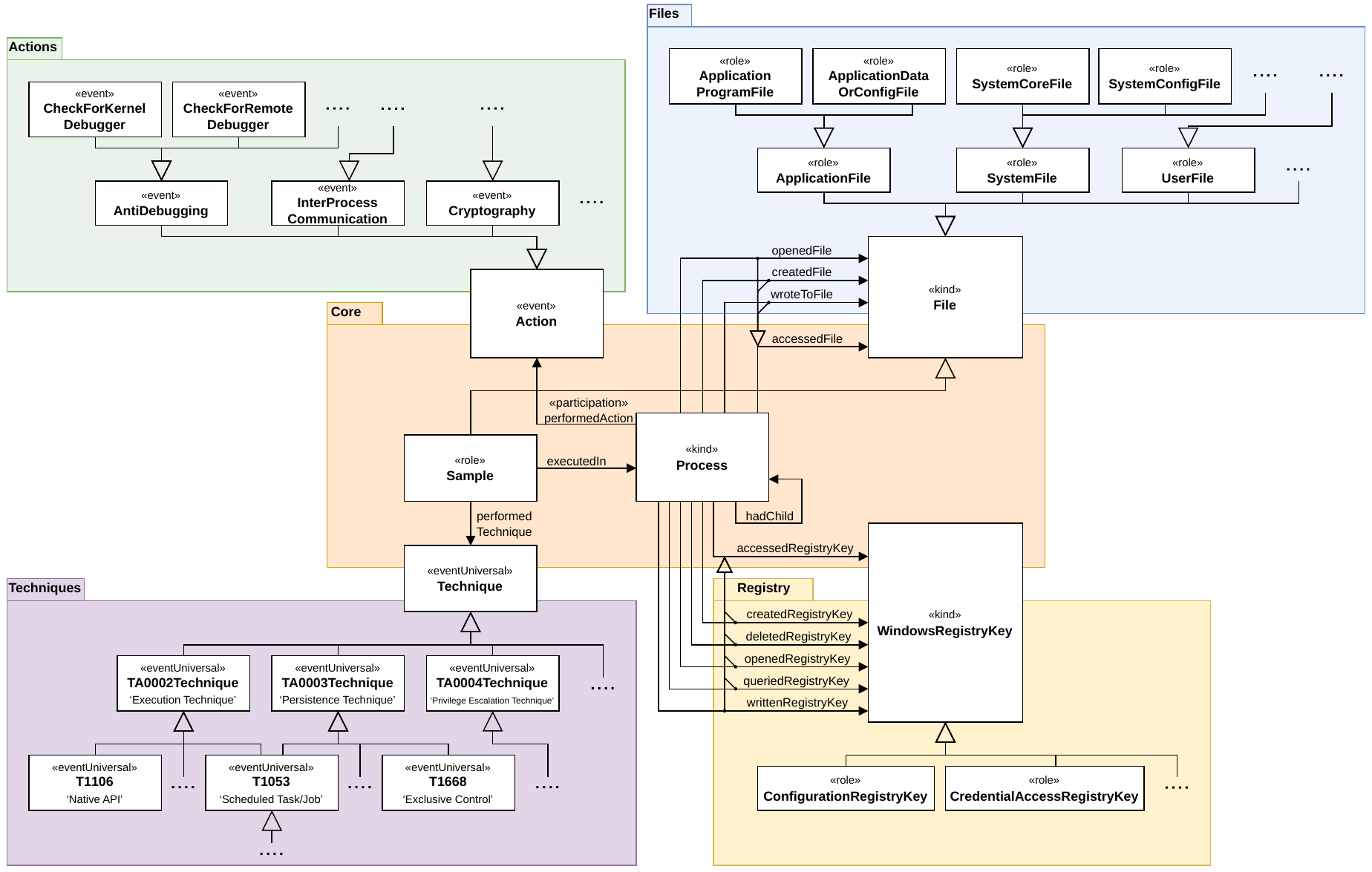}
\caption{MAECO-Lite Ontology}
\label{fig:maeco-lite}
\end{figure}

We call the resulting ontology MAECO-Lite. We view it as a first step towards a possibly larger ontology representing more features available in MAEC (including static features, further observable objects, detailed API calls, etc.). Its simplified (Onto)UML diagram is depicted in Figure~\ref{fig:maeco-lite}.

The ontology models, in support of design goal~1, the runtime behavior of samples through their process trees, API calls \cite{Onoja2026, 27_9936578}, interactions with files and the Windows registry database \cite{Onoja2026,  Amoruso2023User}, and higher-level attacker techniques \cite{Shim2026Techniques}.
It consists of five modules: \emph{Core}, \emph{Actions}, \emph{Files}, \emph{Registry}, and \emph{Techniques}.

The \emph{Core} of the proposed ontology is intentionally kept structurally simple, in line with design goal~2. It comprises six classes: \texttt{Sample}, \texttt{Process}, \texttt{Action}, \texttt{File}, \texttt{WindowsRegistryKey}, and \texttt{Technique}. These are interconnected by five main object properties: \texttt{executedIn}, \texttt{accessedFile}, \texttt{accessedRegistry}, \texttt{performedAction}, and \texttt{performedTechnique}.
This minimal set of classes and relations ensures a lightweight, modular, and easily interpretable structure while capturing the essential aspects of dynamic malware behavior. Additional modules provide specific refinements of some of the core classes and properties.

The core module is centered around the class \texttt{Sample}, roughly corresponding to MAEC's Malware Instance. It is the class of \texttt{File}s subjected to malware analysis. A \texttt{File} is a concrete digital artifact that persists through time, retains identity across analyses
, and bears dispositional properties (e.g., the capability to modify other files). From the UFO perspective, we thus classify the \texttt{Sample} class as a role of substantials (endurants) of the kind \texttt{File}.

Each sample is related to one or more \texttt{Process} instances that executed the sample in a sandbox environment. Processes are further linked to their child processes through the \texttt{hadChild} property, forming a full process tree.
A process is an active runtime context created by the operating system to execute an executable file. A process retains its identity throughout its lifetime, while its memory content and other aspects change. This suggests that processes are endurants; hence, we classify the \texttt{Process} class as a kind in UFO.
Each process is described by three kinds of entities:
a)~the actions it performed, related to it by the \texttt{performedAction} object property,
b)~the files it interacted with, related by the \texttt{accessedFile} property, and
c)~the registry keys it accessed, related by \texttt{accessedRegistry}.

The \emph{Actions} module introduces specialized action subclasses. These abstract the Windows API calls initiated by processes observed during execution. There are 146 classes of actions (derived from the respective MAEC vocabulary) to which API calls can be mapped directly. Each subsumes one of 17 more abstract action classes, such as \texttt{Anti\-Debugging}, \texttt{Inter\-Process\-Communication}, or \texttt{Cryptography} actions, facilitating interpretable abstraction.
The \texttt{Action} class corresponds to the Malware Action type in MAEC, analyzed in Sect.~\ref{sc:onto-analysis-of-maec-stix}. Instances of \texttt{Action} are concrete runtime behaviors observed during dynamic analysis. As such, they are special kinds of events (perdurants) in UFO. Actions unfold in time and correspond to execution steps performed by processes using operating system services or library functions 
and constitute the primary observable evidence of malware activity.
The \texttt{performedAction} object property aligns with the UFO participation relation between an object and an event 
It should be noted that the hierarchy of \texttt{Action} subclasses is structurally identical to the one in the {PE Malware Ontology} \cite{008abs-2403-11669} for static malware analysis. However, here \texttt{Action}s reflect real API calls that were observed during sandbox execution, rather than possible calls statically inferred from the lists of imported functions in executable files. Thus, despite the similarity, MAECO-Lite cannot reuse the \texttt{Action} class hierarchy of {PE Malware Ontology}.

The \emph{Files} module extends the core \texttt{File} class with specialized subclasses and refinements of the \texttt{accessedFile} property to capture file system interactions at a finer level of granularity. Accessed files are classified according to their semantic role within the system, including categories such as \texttt{ApplicationFile}, \texttt{TemporaryFile}, \texttt{UserFile}, and \texttt{SystemFile}. Each of these categories is further refined into more specific subclasses; for example, \texttt{ApplicationFile} is specialized into \texttt{Application\-Data\-Or\-Config\-File} and \texttt{Application\-Program\-File}. In addition, the \texttt{accessedFile} property is refined into more specific relations, such as \texttt{createdFile}, \texttt{openedFile}, and \texttt{writtenToFile}, enabling a more precise representation of how files are involved in malware execution. Ontologically, \texttt{File} is a kind of digital artifacts, and as explained above, its refinements are roles based on relationships with other components of the operating system.

The \emph{Registry} module similarly extends the \texttt{WindowsRegistryKey} class and refines the \texttt{accessedRegistry} property to capture registry-related interactions in greater detail. Registry keys are classified according to the role they can play in malware behavior, rather than their structural properties (e.g., path). 
This gives rise to classes such as \texttt{PersistenceRegistryKey}, \texttt{ConfigurationRegistryKey}, \texttt{CredentialAccess}\-\texttt{RegistryKey}, \texttt{DefenseEvasionRegistryKey}, etc.
The \texttt{accessed}\-\texttt{Registry} property is specialized into more fine-grained relations, including \texttt{created}\-\texttt{Registry\-Key}, \texttt{deleted\-Registry\-Key}, \texttt{opened\-Registry\-Key}, \texttt{queried\-Registry\-Key}, and \texttt{written}\-\texttt{To}\-\texttt{Registry\-Key}, enabling precise description of how registry keys are manipulated during malware execution.
Ontologically, \texttt{Windows\-Registry\-Key} 
is a kind in UFO, classifying system resources that bear a distinct identity. Its refinements, given the semantics explained above, are roles.

The \emph{Techniques} module extends the ontology with 691 subclasses of \texttt{Technique} derived from MITRE ATT\&CK techniques, e.g., \texttt{T1053} labeled (abuse of) `Scheduled Task/Job' \cite{mitre-attack}. \texttt{Technique}s are global behavioral patterns composed of multiple actions that may span several processes, and these actions are ultimately manifestations of a \texttt{Sample}'s dispositions. \texttt{Technique}s are thus linked to the \texttt{Sample} via the \texttt{performed}\-\texttt{Technique} property. Additionally, there are 14 more abstract classes (e.g., \texttt{TA0002Technique}, labeled `Execution Technique') which subsume techniques that may contribute to a particular ATT\&CK tactic, a high-level adversarial goal. A technique may contribute to multiple tactics (e.g., Scheduled Task/Job abuse contributes to Execution and Persistence). Tactic-based classes serve as interpretable abstractions of lower-level techniques. Ontologically, a \texttt{Technique} is an event universal (perdurant type). At the moment, the ontology is not modeling individual \texttt{Technique} occurrences and their composition from \texttt{Action}s observed during the analysis. This information would be of no additional benefit in the intended usage scenarios. However, an module providing this capability may be added in the future.

\section{Initial Evaluation}

\begin{table}[tp!]
    \vspace*{-1.5ex}\caption{Results from baseline MAEC ontology.}
    \label{tab:results_maeco}
    \scriptsize\centering\setlength{\tabcolsep}{0.5em}%
    \begin{tabular}{ l c c c c c }
        \toprule
        {Algorithm} & {Accuracy} & {Precision} & {Recall} & {FP Rate} & {F1} \\
        \midrule
        {OCEL}  & 0.49 $\pm$ 0.00 & 0.49 $\pm$  0.00& 1.00 $\pm$ 0.00 & 1.00 $\pm$ 0.00 & 0.66 $\pm$ 0.00 \\
        {CELOE}  & 0.49 $\pm$ 0.00 & 0.49 $\pm$  0.00& 1.00 $\pm$ 0.00 & 1.00 $\pm$ 0.00 & 0.66 $\pm$ 0.00 \\
        {PARCEL}  & 0.50 $\pm$ 0.00 & 0.00 $\pm$ 0.00  & 0.00 $\pm$ 0.00  & 0.00 $\pm$ 0.00 & 0.00 $\pm$ 0.00  \\
    {SPACEL}  & 0.50 $\pm$ 0.00  & 0.00 $\pm$ 0.00  & 0.00 $\pm$ 0.00  & 0.00 $\pm$ 0.00  & 0.00  $\pm$ 0.00 \\
        \bottomrule
    \end{tabular}
\end{table}

\begin{table}[tp!]
    \vspace*{-1.5ex}\caption{Results from MAECO-Lite ontology.}
    \label{tab:results_maeco_lite}
    \scriptsize\centering\setlength{\tabcolsep}{0.5em}%
    \begin{tabular}{ l c c c c c }
        \toprule
        {Algorithm} & {Accuracy} & {Precision} & {Recall} & {FP Rate} & {F1} \\
        \midrule
        {OCEL}  & 0.79 $\pm$ 0.03 & 0.77 $\pm$  0.03& 0.83 $\pm$ 0.03 & 0.18 $\pm$ 0.03 & 0.84 $\pm$ 0.03 \\
        {CELOE}  & 0.74$\pm$ 0.01  & 0.68 $\pm$ 0.01 & 0.89 $\pm$ 0.01  & 0.41 $\pm$ 0.03  & 0.77 $\pm$ 0.01  \\
        {PARCEL}  & 0.80 $\pm$ 0.01 & 0.86 $\pm$ 0.02  & 0.73 $\pm$ 0.01  & 0.11 $\pm$ 0.02 & 0.79 $\pm$ 0.01  \\
    {SPACEL}  & 0.71 $\pm$ 0.03  & 0.90 $\pm$ 0.02  & 0.48 $\pm$ 0.06  & 0.05 $\pm$ 0.01  & 0.62  $\pm$ 0.06 \\
        \bottomrule
    \end{tabular}
\end{table}

To assess the suitability of both ontologies for concept learning, we conducted an initial evaluation using the DL-Learner framework with several state-of-the-art algorithms, namely OCEL, CELOE, PARCEL and SPACEL \cite{lehmann2009dl}. Both ontologies under evaluation differ in their construction. MAEC ontology is a baseline ontology naively derived from the structure of MAEC JSON report files, where the original flat hierarchy of the report attributes is mirrored in the ontology. MAECO-Lite, on the other hand, is an ontologically aligned reformulation that introduces class hierarchy and semantically motivated relations. Both ontologies contain 1000 samples (500 malware and 500 benign). The experiments used 5-fold cross-validation with an 80/20 train--test split, with each fold allocated 5 minutes of user time\footnote{The experiments were run on a system with an 18-core Intel Core i9-10980XE processor, 256 GB of RAM, running Debian 5.10-140.}. Note that the storage difference between the two ontologies reflects not the ontology schema itself, but the ontologically aligned instance data in RDF: the MAEC baseline RDF dataset occupies approximately 90~MB on disk, compared to only 14~MB for MAECO-Lite. Both ontologies were successfully loaded and evaluated under identical experimental conditions. Results for 
baseline MAEC and MAECO-Lite are reported in Tables~\ref{tab:results_maeco} and~\ref{tab:results_maeco_lite}, respectively. 

On MAECO-Lite, all algorithms yielded meaningful results with F1 scores ranging from 62\% to 84\%. In contrast, experiments at the MAEC baseline yielded near-trivial results: OCEL and CELOE both returned $\exists\,\mathtt{triggeredSignature}.\top$ as the learned class expression, effectively classifying every instance that triggered at least one signature as malware (recall of 1.00, precision of 0.49 reflecting the class balance), without exploiting any deeper relational structure. PARCEL and SPACEL defaulted to predicting all instances as negative (precision, recall, and F1 of 0.00). These findings suggest that while MAECO-Lite provides a tractable learning environment with sufficient representational coverage for effective concept learning, the increased expressiveness of the MAEC baseline ontology poses a significant challenge for the evaluated algorithms.

We can see two examples of class expressions below. \begin{align}
    \label{ocel}
    &\text{\small$\begin{aligned}
        &\tt\exists executedIn . \exists performedAction . \top \\[-\jot]
        &\tt\sqcap \atleast{2} \: performedTechnique . \texttt{`Defense Evasion Technique'}
    \end{aligned}$}
    \\
    \label{parcel}
    &\text{\small$\begin{aligned}
        &\tt\exists performedTechnique . \texttt{`Process Hollowing'} \\[-\jot]
        &\tt\sqcap \exists performedTechnique . \texttt{`Inter-Process Communication'}
    \end{aligned}$}
\end{align}
The first class expression \eqref{ocel}, derived by OCEL, characterizes malware that executes in an environment where at least one action was observed and employs two or more \textit{Defense Evasion} techniques (these include methods used by malware to avoid detection, hinder its analysis, and generally conceal its malicious activity from security analysts). The second class expression \eqref{parcel}, derived by PARCEL, characterizes malware that combines the \textit{process hollowing} technique (hiding malicious code inside a legitimate process), while inter process communication enables the injected code to interact with other processes.

\section{Related Work}
Several ontology-based approaches have been proposed for representing malware knowledge, differing in scope, level of formalization, and support for dynamic behavior.
Ding et al. \cite{003DingWZ19} developed a knowledge base about malware instances and families, leveraging behavioral features to group malware samples and emplyoing  pattern mining techniques for classification. It is primarily oriented toward data organization and knowledge storage, lacking a formally grounded ontology with explicit semantic commitments. 
Behavior-based approaches, e.g.\ Chowdhury et al. \cite{006ChowdhuryB22}, Chiang et al. \cite{chiang2010}, and Grégio et al. \cite{005GrecioBNAGJ}, 
typically represent behavior as sets of features, patterns, or rules derived from system activities. While effective for identifying recurring patterns and supporting detection tasks, they generally lack a clear ontological distinction between behavioral types, their realizations at runtime, and the entities involved. Consequently, behaviors are often conflated with observations or attributes, leading to semantic ambiguities and limiting their suitability for fine-grained reasoning about dynamic malware execution.

The PE Malware Ontology \cite{008abs-2403-11669} provides a reusable semantic schema for representing features of Portable Executable files, leveraging the MAEC standard for describing possible malware actions. The ontology is predominantly centered on static analysis features, such as file structure, metadata, and imported functions. It is thus less suitable for capturing runtime behavior and inherits limitations associated with static analysis, including susceptibility to obfuscation \cite{MOLINACORONADO2025104094} and evasion techniques \cite{GENG2024103595}.
Balogh and Galko \cite{007BaloghG23} proposed integrating static and dynamic analysis into a unified MAEC-based representation. They demonstrate the feasibility of aggregating heterogeneous data sources into a single model. 
However, the representation is data-driven, lacking explicit ontological distinctions for modeling dynamic phenomena. In particular, actions and behaviors are not formally treated as temporally extended events involving participating entities, which limits the expressiveness and semantic precision of the model. 

Overall, existing approaches either emphasize static characteristics or incorporate dynamic features without a principled ontological foundation. In many cases, runtime aspects such as actions, behaviors, and techniques are modeled as attributes or labels of malware artifacts rather than as distinct event-based entities. This limits their ability to capture the temporal structure of malware execution and constrains reasoning capabilities.
In contrast, our work adopts a foundational ontology standard to provide a principled grounding for malware modeling. We explicitly distinguish between malware artifacts (endurants), their capabilities (dispositions), and runtime behaviors (perdurants), and model actions as temporally extended events involving well-defined participants. This ontologically grounded approach improves conceptual clarity, supports more expressive reasoning, and enables a semantically consistent integration of static and dynamic malware analysis within a unified framework.
\section{Conclusion and Future Work}
This paper addresses the challenge of representing dynamic malware behavior in a semantically principled and computationally useful way. Although widely adopted standards such as MAEC and STIX provide rich vocabularies for malware analysis, they often blur important ontological distinctions, particularly between enduring artifacts and runtime events. To clarify these issues, we conducted a foundational analysis of MAEC constructs and selected STIX Cyber Observable Objects using the Unified Foundational Ontology (UFO). The analysis revealed several ontological mismatches, motivating the design of a lightweight ontology, MAECO-Lite, tailored to represent dynamic malware execution. The ontology adopts a modular structure centered on malware samples, processes, actions, system artifacts, and attacker techniques, while maintaining a clear separation between substantials, events, and dispositions.

An initial evaluation using description logic concept learning algorithms shows that the simplified ontology significantly improves learning performance compared to a naive baseline. Future work will extend the ontology with additional runtime artifacts and evaluate it on larger dynamic malware datasets.

\section*{Declaration on Generative AI}
During the preparation of this work, the authors used \emph{Copilot} and \emph{Writefull} for grammar and spelling checks, improving writing style. The authors subsequently reviewed and edited the content as necessary and take full responsibility for the publication's content.

\bibliographystyle{vancouver}
\bibliography{references}
\end{document}

%% file: tab-maec-stix-ufo-mapping.tex
\renewcommand{\arraystretch}{1.25}
\begin{tabular}{|
l|
l|
>{\raggedright\arraybackslash}p{7.75cm}|
}
\hline
\textbf{MAEC / STIX Concept} & \textbf{UFO Category} & \textbf{ Interpretation Note} \\
\hline

\multicolumn{3}{|l|}{\emph{MAEC Types and Top-Level Objects}} \\
\hline
Malware Instance  & Role & Context-dependent role played by a File subjected to malware analysis within the analysis and operational context.  \\
\hline
Process Tree Node & \makecell[lt]{Processual Role\\ + Relation} & Process’s processual role as actions initiator, result; ordinal position w.r.t. to other processes.  \\
\hline
Malware Family & Universal & Analyst-defined classification grouping malware instances; membership is not intrinsic \\
\hline
\makecell[lt]{Capability / Behavior / \\ Action (statically inferred)} & Disposition & Potential of malware to perform actions (e.g., keylogging), manifested through execution events \\
\hline
Behavior (occurrence) & Complex Event  & Complex event composed of multiple actions representing the realization of a malware capability \\
\hline
Action (occurrence) & Event & Concrete execution event representing a system-level operation (e.g., file access, process creation) \\
\hline
API Call & Atomic Event & Fine-grained execution event representing invocation of a system or library function \\
\hline
Static Feature / Binary Obfuscation & Intrinsic Moments & Intrinsic structural property (e.g., hash, size, imports) \\
\hline
\makecell[lt]{Dynamic Features /\\ Malware Development Environment} & Relator & Structured information artifact describing observed and inferred aspects of malware execution, including events and related entities \\
\hline
\makecell[tl] {Signature Metadata / Analysis Metadata / \\ Field Data }  &  Object & Description of analysis results, detection context, and provenance \\
\hline
Name & Quality  & Symbolic label assigned by analysts or systems; conventional and not intrinsic \\
\hline

\multicolumn{3}{|l|}{\emph{STIX 2.1 Cyber Observable Objects (SCOs)}} \\
\hline
Observable Object & Role Mixin  & Evidence-bearing digital entities observed in a system, used to describe the state and artifacts involved in cybersecurity operation. \\
\hline
\makecell[tl]{File / Directory / Artifact / Process / \\ Mutex / WindowsRegistryKey / Software / \\ X.509 Certificate / Email Message / \\ User Account / Autonomous System} & Object  & Identity-bearing digital artifact that persists through time and can play roles in certain contexts. \\
\hline
Network Traffic & Event & Temporally extended communication event involving data exchange between network entities \\
\hline
IPv4 / IPv6 / MAC Address & Quality & Identifying quality inhering in network interfaces or network-capable entities \\
\hline
 Domain Name / URL / Email Address & Object & Symbolic identifier governed by naming conventions and resolution systems.  \\
\hline

\end{tabular}%